# Robust quantitative single-exposure laser speckle imaging with true flow speckle contrast in the temporal and spatial domains

CHENGE WANG,[1,#] ZILI CAO,[1,#] XIN JIN,[1] WEIHAO LIN,[1] YANG ZHENG,[1] BIXIN ZENG,[1] AND M. XU[1,2,*]

[1] *Institute of Lasers and Biophotonics, School of Biomedical Engineering, Wenzhou Medical University, Wenzhou, Zhejiang, China 325035*
[2] *Department of Physics and Astronomy, Hunter College, The City University of New York, 695 Park Avenue, New York, NY 10065*
#: *These authors equally contributed to the work.*
*\*minxu@hunter.cuny.edu*

**Abstract:** A systematic and robust laser speckle contrast imaging (LSCI) method and procedure is presented, covering the LSCI system calibration, static scattering removal, and measurement noise estimation and correction to obtain a true flow speckle contrast $K_f^2$ and the flow speed from single-exposure LSCI measurements. We advocate to use $K^2$ as the speckle contrast instead of the conventional contrast K as the former relates simply to the flow velocity and is with additive noise alone. We demonstrate the efficacy of the proposed true flow speckle contrast by imaging phantom flow at varying speeds, showing that (1) the proposed recipe greatly enhances the linear sensitivity of the flow index (inverse decorrelation time) and the linearity covers the full span of flow speeds from 0 mm/s to 40 mm/s; and (2) the true flow speed can be recovered regardless of the overlying static scattering layers and the type of speckle statistics (temporal or spatial). The fundamental difference between the apparent temporal and spatial speckle contrasts is further revealed. The flow index recovered in the spatial domain is much more susceptible to static scattering and exhibit a shorter linearity range than that obtained in the temporal domain. The proposed LSCI analysis framework paves the way to estimate the true flow speed in the wide array of laser speckle contrast imaging applications.

## 1. Introduction

When coherent light interacts with a turbid medium, the interference between the outgoing waves produces grainy speckle patterns which encode the phase fluctuation of all rays (random phasors) reaching a point. The contrast of laser speckles reduces with the motion of scatterers inside the turbid medium. Laser speckle contrast hence can be used to infer the dynamic property of the medium. Laser speckle contrast imaging (LSCI, see recent reviews [1-4]) has now been widely used in monitoring blood flow in brain, skin, retina, arthrosis and etc due to advantages including simplicity, high spatial and temporal resolution, and large field of view without scanning [5-8].

Although LSCI has a wide range of applications and a long history, the recovery of absolute flow velocity from LSCI measurements remains a challenge, especially when the measurement is compounded by static scattering and noise. For static scattering in laser speckle imaging, Li et al. showed that the static scattering effect can be partially suppressed by using the temporal rather than spatial contrast analysis of laser speckles [9] as the static scattering is an invariant quantity with time. Zakhraov et al. [10, 11] presented a data processing scheme to correctly separate dynamic and static components within the speckle contrast based on their different decorrelation behaviour across speckle patterns captured at consecutive times. Dunn et al. [6,

12-14] later demonstrated a multi-exposure laser speckle contrast imaging method, which quantifies and eliminates the influence of static scattering from speckle contrasts measured under different exposure times using a laser speckle contrast model. For LSCI measurement noise, the correction of the variance of the shot noise and sensor dark currents were found to be crucial to estimate the true speckle contrast [15, 16]. Yuan et al. [16] increased the signal-to-noise ratio (SNR) of LSCI with noise correction to detect small blood flow changes caused by brain activity. A systematic study and recommended practical recipe to obtain true flow velocity from LSCI measurements addressing both static scattering and measurement noise is, however, still lacking.

In this article, we analysed laser speckle flow imaging from the first principle and provided a complete procedure covering the LSCI system calibration, static scattering removal, and measurement noise estimation and correction to obtain a genuine flow speckle contrast and the flow speed from single-exposure LSCI measurements. We demonstrated the power of our LSCI analysis recipe by imaging phantom flow at varying speeds. Experimental results show that our procedure greatly enhances the linear sensitivity of the flow index (defined as the inverse decorrelation time) and the linearity covers the full span of flow speeds from 0 mm/s to 40 mm/s. The true flow speed is recovered regardless of the overlying static scattering layers and the type of statistics (temporal or spatial). The proposed LSCI analysis framework hence paves the way to estimate the true flow speed in the wide array of laser speckle contrast imaging applications.

## 2. Theory and Data Analysis

### 2.1 Theoretical basis

The spatial intensity distribution of the speckle pattern fluctuates with the motion of the scattering particles under the illumination of coherent light. The recorded pattern by a camera is the integration of all instantaneous speckles over the exposure time. The faster the scattering particles move, the more blurred the recorded pattern becomes. The degree of blurring is quantified by the contrast [17] given by

$$K = \frac{\sqrt{\sum (I - \bar{I})^2}}{\bar{I}} \tag{1}$$

where $\bar{I}$ represents the mean of light intensity $I$ over a small region (spatial contrast) or over a short durance of time (temporal contrast). For "fully developed" static speckles, the spatial contrast K equals to 1.

We will assume the scattered electric field containing both dynamic and static components

$$E(t)e^{-i\omega t} = [E_f(t) + E_s]e^{-i\omega t} \tag{2}$$

with ω being the angular frequency of light. The dynamic component consists of photons which have at least been scattered by moving scatterers (flow) once and the static component consists of photons being scattered by static scatterers alone.

The electric field temporal autocorrelation function can be written as

$$\langle E(t)E^*(t+\tau) \rangle = G_1(\tau) + I_s \tag{3}$$

where $\langle \ \rangle$ means average over $t$, $G_1(\tau) = \langle E_f(t)E_f^*(t+\tau) \rangle$ is the electric field temporal autocorrelation function related to flow, and $I_s = |E_s|^2$.

In practice only light intensity fluctuation signals can be recorded. The intensity autocorrelation function is defined as $G_2(\tau) = \langle I(t)I(t+\tau) \rangle$ where $I(t) = I_f(t) + I_s$ and $I_f(t) = |E_f(t)|^2$ assuming the dynamic and static electric fields are uncorrelated, i.e., $\langle E_f(t)E_s^* \rangle = 0$. In terms of the normalized dynamic electric field and full intensity autocorrelations $g_1(\tau) = G_1(\tau)/\overline{I_f}$ and $g_2(\tau) = G_2(\tau)/G_2(0)$, the full Siegert relation [18, 19] expresses

$$g_2(\tau) = 1 + \frac{\beta}{(\overline{I_f} + I_s)^2}\left[\overline{I_f}^2 |g_1(\tau)|^2 + 2\overline{I_f}I_s |g_1(\tau)|\right] \tag{4}$$

where $\overline{I_f} = \langle I_f(t) \rangle$, $\beta \leq 1$ is a parameter that accounts for the reduction in the measured contrast due to averaging (by the detector) over uncorrelated speckles. Note $g_1$ is real and non-negative [19].

The speckle contrast under an exposure time $T$, expressed as $K^2 = T^{-2}\left\langle \int_0^T \int_0^T I(t_1)I(t_2)dt_1 dt_2 \right\rangle / \langle I \rangle^2 - 1$, reduces now to

$$K^2 = \rho^2 \frac{2\beta}{T}\int_0^T (1-\frac{\tau}{T})|g_1(\tau)|^2 d\tau + 2\rho(1-\rho)\frac{2\beta}{T}\int_0^T (1-\frac{\tau}{T})|g_1(\tau)|d\tau \tag{5}$$

where the average intensity $\langle I \rangle = \overline{I_f} + I_s$ and the dynamic fraction $\rho = \overline{I_f}/(\overline{I_f} + I_s)$. We will use $K^2$ as the speckle contrast instead of the conventional contrast $K$ as the former relates simply to the flow velocity and is with additive noise alone.

Equation (5) is the theoretical temporal contrast from the random process taken by the photons migrating through a turbid medium. Some complexities arise when evaluating $K^2$ from measurement data. First the measurement noise (of zero mean) introduces an extra variance term $\kappa_{noise}^2$. Second when using spatial ensemble average for the evaluation of the contrast rather than temporal statistics, the extra terms appearing in $K^2$ will be $\kappa_{noise}^2 + \kappa_{ne}^2$ with $\kappa_{ne}^2$ being the non-ergodic contribution from the static field. This motivates us to introduce the dynamic (flow) contrast $K_f^2 = \langle I_f^2 \rangle / \langle I_f \rangle^2 - 1$ defined in terms of the dynamic component alone. The measured speckle contrast can be expressed as

$$K^2 = \rho^2 K_f^2 + \kappa_{noise}^2 \tag{6}$$

in the temporal domain and

$$K^2 = \rho^2 K_f^2 + \kappa_{noise}^2 + \kappa_{ne}^2 \tag{7}$$

in the spatial domain. The value of $\kappa_{noise}^2$ and $\kappa_{ne}^2$ can be evaluated from calibration and measurement data as shown later. The dynamic contrast remains the same when evaluated in either the temporal or spatial domain.

A velocity distribution model for the moving particles is needed to relate $K_f^2$ to the flow speed. With the commonly used Lorentz velocity distribution model, the dynamic electric field autocorrelation function can be written as $g_1(\tau) = \exp(-t/\tau_c)$ [12], yielding

$$K_f^2 = \beta \frac{\exp(-2x)-1+2x}{2x^2} + 4\beta(\rho^{-1}-1)\frac{\exp(-x)-1+x}{x^2}$$

$$= \beta(2\rho^{-1}-1) - \frac{2}{3}\beta\rho^{-1}x + \frac{1}{6}\beta(\rho^{-1}+1)x^2, \quad x \ll 1 \tag{8}$$

$$= \beta(4\rho^{-1}-3)\frac{1}{x} + \beta(\frac{7}{2}-4\rho^{-1})\frac{1}{x^2}, \quad x \gg 1$$

where $x = T/\tau_c$ and $\tau_c$ is the decorrelation time. The inverse decorrelation time increases with the flow speed and can be regarded as the flow index.

In the next section, we will examine system calibration, sample measurement and data analysis to provide a complete procedure for static scattering removal, and measurement noise estimation and correction to obtain a true flow speckle contrast and the flow speed from single-exposure LSCI measurements.

## 2.2 System calibration, measurement, and data analysis

Let's consider a set of speckle images $I_i$ ($1 \leq i \leq N$) at time $t_i = i\Delta t$ with an exposure time $T$, i.e.,

$$I_i(x, y) = I_s(x, y) + I_{fi}(x, y) + n_i(x, y) \tag{9}$$

Here the recorded image consists of the static component $I_s$, the dynamic component $I_f$ and the noise $n$. The noise [15] presented in the measurement is mainly comprised of the dark counts $n_d$ and the shot noise $n_s$, i.e., $n_i(x, y) = n_{di}(x, y) + n_{si}(x, y)$.

The dark counts $n_d$ and the variance of dark counts $Var(n_d) = \langle n_d^2 \rangle - \langle n_d \rangle^2$ can be easily acquired by taking multiple dark frames at the same exposure time and camera gain (with all light off). One could use the temporal average to get the dark count and its variance pixel by pixel when the number of the dark frames is large ($\geq 50$) or use the spatial average over a sliding $N_P \times N_P$ (typical $N_P = 7$) pixel window otherwise. If the behavior of dark counts is assumed uniform across the whole sensor frame, the mean and the variance of the dark counts are given by further averaging over the whole sensor frame. In many cameras, the recorded intensity has been pre-subtracted by certain base. In this case, $\langle n_d \rangle$ should be estimated by the median value and the variance $Var(n_d) = 2\langle n_d'^2 \rangle$ where $n_d' = n_d - \langle n_d \rangle$ with the negative values of $n_d'$ replaced by zeros.

We would always subtract the dark counts from $I_i$ before further analysis. After this pre-processing, the speckle image becomes $I_i^c(x, y) = I_i(x, y) - \langle n_d(x, y) \rangle$ and the noise term is replaced by $n_i^c(x, y) = n_i(x, y) - \langle n_d(x, y) \rangle$. The noise satisfies $\langle n_i^c \rangle = 0$ and $Var(n_i^c) = Var(n_{si}) + Var(n_d)$. The shot noise obeys a Poisson distribution with the mean $\langle n_{si} \rangle = 0$ and a variance equal to $\langle I_i^c(x, y) \rangle / \gamma$ as the camera converts the photoelectrons to digital counts where $\gamma$ is the analog to the digital conversion factor[20]. The $\gamma$ factor is typically the same across the sensor frame and thus is obtained by further averaging over the sensor frame. Under such a shot noise model,

$$Var(n_i^c) = \alpha \langle I_i^c(x, y) \rangle^2 + \langle I_i^c(x, y) \rangle / \gamma + Var(n_d) \tag{10}$$

where an extra quadratic term in $\langle I_i^c(x,y) \rangle$ is added to account for other noise sources such as the laser fluctuations.

The temporal and spatial averages of the sequence of single-exposure speckle images then satisfy:

$$\langle I_i^c(x,y) \rangle_i = I_s(x,y) + \langle I_{fi}(x,y) \rangle_i \tag{11}$$

$$\langle I_i^c(x,y) I_j^c(x,y) \rangle_i = I_s^2(x,y) + \langle I_{fi}(x,y) I_{fj}(x,y) \rangle_i + 2 I_s(x,y) \langle I_{fi}(x,y) \rangle_i \tag{12}$$

and

$$\langle I_i^{c2}(x,y) \rangle_i = I_s^2(x,y) + \langle I_{fi}^2(x,y) \rangle_i + \langle n_i^{c2}(x,y) \rangle_i + 2 I_s(x,y) \langle I_{fi}(x,y) \rangle_i \tag{13}$$

with temporal statistics and

$$\langle I_i^c(x,y) \rangle_{xy} = \langle I_s(x,y) \rangle_{xy} + \langle I_{fi}(x,y) \rangle_{xy} \tag{14}$$

$$\langle I_i^c(x,y) I_j^c(x,y) \rangle_{xy} = \langle I_s^2(x,y) \rangle_{xy} + \langle I_{fi}(x,y) I_{fj}(x,y) \rangle_{xy} + 2 \langle I_s(x,y) \rangle_{xy} \langle I_{fi}(x,y) \rangle_{xy} \tag{15}$$

and

$$\langle I_i^2(x,y) \rangle_{xy} = \langle I_s^2(x,y) \rangle_{xy} + \langle I_{fi}^2(x,y) \rangle_{xy} + 2 \langle I_s(x,y) \rangle_{xy} \langle I_{fi}(x,y) \rangle_{xy} + \langle n_i^{c2}(x,y) \rangle_{xy} \tag{16}$$

with spatial statistics. Here $j \equiv i + \Delta$ with $\Delta \neq 0$, $\langle \ \rangle_i$ means averaging over the $N$ temporal instances and $\langle \ \rangle_{xy}$ means the spatial average over a sliding $N_P \times N_P$ pixel window. We have used the fact $\langle I_{fi}(x,y) \rangle_{xy} = \langle I_{fj}(x,y) \rangle_{xy}$ due to the ergodic nature of the dynamic component. When the time difference satisfies $t_j - (t_i + T) \gg \tau_c$, the complete decorrelation between the dynamic component measured at two different times ensures the important identities $\langle I_{fi}(x,y) I_{f,i+\Delta}(x,y) \rangle_i = \langle I_{fi}(x,y) \rangle_i \langle I_{f,i+\Delta}(x,y) \rangle_i$ and $\langle I_{fi}(x,y) I_{fj}(x,y) \rangle_{xy} = \langle I_{fi}(x,y) \rangle_{xy} \langle I_{fj}(x,y) \rangle_{xy}$ as $E_{fi}(x,y)$ is a zero-mean Gaussian variable. These identities could serve as the data consistency check.

One important consequence is that for any two speckle images $I_i$ and $I_j$ taken at times satisfying $t_j - (t_i + T) \gg \tau_c$, the following holds

$$\frac{\langle I_i^c(x,y) I_j^c(x,y) \rangle_{xy} - \langle I_i^c(x,y) \rangle_{xy} \langle I_j^c(x,y) \rangle_{xy}}{\langle I_i^c(x,y) \rangle_{xy}^2} = \beta(N_p)(1-\rho)^2 \tag{17}$$

where the coherence factor of the imaging system is defined as

$$\beta(N_p) = \frac{\langle I_s^2(x,y) \rangle_{xy}}{\langle I_s(x,y) \rangle_{xy}^2} - 1 \tag{18}$$

associated with spatial averaging over the sliding window of size $N_P$. Equation (17) is also correct for a pure static scattering sample producing fully developed speckles ($\rho = 0$) as long as $j \neq i$.

### 2.2.1 System calibration

As stated earlier, the dark counts $n_d$ and the variance of dark counts $Var(n_d)$ is first acquired by taking multiple dark frames at the identical experimental condition (the same exposure time, camera gain and etc) with all light off. The other system parameters (the coherence factor $\beta$ and the behavior of $Var(n^c)$ vs $\langle I^c \rangle$) can be directly evaluated from a set of fully developed speckle images taken on a pure static scattering sample such as a reflection standard. Indeed, according to Eq. (17), we have

$$\beta(N_p) = \frac{\langle I_i^c(x,y) I_{i+\Delta}^c(x,y) \rangle_{xy}}{\langle I_i^c(x,y) \rangle_{xy} \langle I_{i+\Delta}^c(x,y) \rangle_{xy}} - 1 \tag{19}$$

for $\Delta \neq 0$ in this case. Here the spatial average should use the largest window size (full image if possible) due to the reason discussed in [21] for the temporal speckle contrast. For the spatial speckle contrast $N_p$ in Eq. (19) should take the same value used for the contrast calculation. The $\beta$ factor is typically the same across the sensor frame and thus obtained by further averaging over the sensor frame.

The noise variance $Var(n^c)$ associated with a particular $\langle I^c \rangle$ can be found as

$$Var(n^c) = \langle I_i^{c2}(x,y) \rangle_i - \langle I_i^c(x,y) \rangle_i^2 \tag{20}$$

or

$$Var(n^c) = \sqrt{\langle I_i^{c2}(x,y) \rangle_{xy} \langle I_{i+\Delta}^{c2}(x,y) \rangle_{xy}} - \langle I_i^c(x,y) I_{i+\Delta}^c(x,y) \rangle_{xy} \tag{21}$$

with the temporal or spatial statistics, respectively. Multiple sets of such speckle images under the identical experimental condition and varying incident intensities are measured to cover the full range of $\langle I^c \rangle$. By fitting to Eq. (10), the system noise behavior is then characterized.

We note the above results on $\beta$ and $Var(n^c)$ should be independent of the $\Delta \neq 0$. In reality a slight dependence on $\Delta$ can be observed due to the inevitable system instability. In this case, the correct values of $\beta$ and $Var(n^c)$ are obtained by extrapolating to $\Delta = T/2\Delta t$.

### 2.2.2 Sample measurement and data processing

The sample containing both dynamic and static scatterers are then imaged under the identical experimental condition to yield a new set of dark current removed speckle images $I_i^c(x,y)$ ($1 \leq i \leq N$).

The dynamic fraction $\rho$ can be determined using Eq. (17) as

$$(1-\rho)^2 = \frac{1}{\beta(N_p)} \frac{\langle I_i^c(x,y) I_{i+\Delta}^c(x,y) \rangle_{xy}}{\langle I_i^c(x,y) \rangle_{xy}^2} - 1 \tag{22}$$

for $t_{i+\Delta} - (t_i + T) \gg \tau_c$.

The noise level can also be estimated directly from the set of speckle images via

$$\kappa_{noise}^2 = \frac{\alpha \langle I_i^c(x,y) \rangle_i^2 + \langle I_i^c(x,y) \rangle_i / \gamma + Var(n_d)}{\langle I_i^c(x,y) \rangle_i^2} \quad (23)$$

in the temporal domain or

$$\kappa_{noise}^2 = \frac{\alpha \langle I_i^c(x,y) \rangle_{xy}^2 + \langle I_i^c(x,y) \rangle_{xy} / \gamma + Var(n_d)}{\langle I_i^c(x,y) \rangle_{xy}^2} \quad (24)$$

in the spatial domain (they are equal by ergodicity).

Using Eq. (13) and (16), the temporal speckle contrast is then

$$K^2 = \frac{\langle I_i^{c2}(x,y) \rangle_i}{\langle I_i^c(x,y) \rangle_i^2} - 1 = \rho^2 K_f^2 + \kappa_{noise}^2 \quad (25)$$

and the spatial speckle contrast is then

$$K^2 = \frac{\langle I_i^{c2}(x,y) \rangle_{xy}}{\langle I_i^c(x,y) \rangle_{xy}^2} - 1 = \rho^2 K_f + \beta(N_p)(1-\rho)^2 + \kappa_{noise}^2 \quad (26)$$

from which we can identify the non-ergodic contribution from the static field to be

$$\kappa_{ne}^2 = \beta(N_p)(1-\rho)^2 \quad (27)$$

Finally, the velocity information of the sample can be obtained by solving the flow contrast $K_f^2$ and fitting to Eq. (8) to obtain the decorrelation time $\tau_c$ and the flow index.

Figure 1 outlines the complete procedure of system calibration, sample measurement, noise correction, and static scattering removal for robust quantitative single-exposure laser speckle imaging.

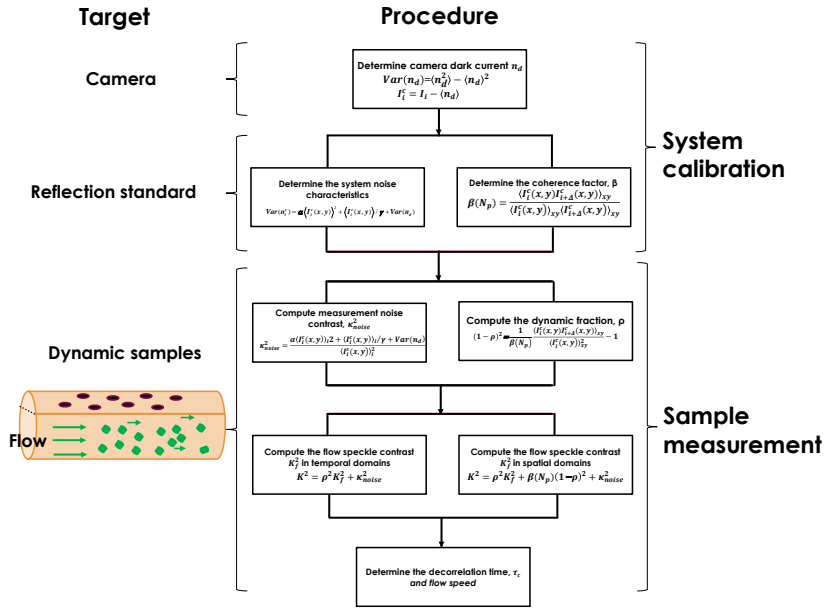

**Fig. 1** Experimental and data analysis framework for robust quantitative single-exposure laser speckle imaging.

System calibration first determines the camera dark current $\langle n_d \rangle$, the coherent factor $\beta$, and the noise parameters α and $\gamma$ from measuring fully developed speckles produced by a pure static reflection standard. The true flow contrast $K_f^2$ of dynamic samples are afterwards obtained by removing the static scattering and correcting the measurement noise from the measured temporal or spatial speckle contrasts. The flow decorrelation time and speed are then be determined from $K_f^2$ with a proper flow velocity model such as Eq. (8).

## 3. Results

### 3.1 Experimental setup

Figure 2 shows the schematic diagram of the experimental setup. Light from a DPSS red laser (LSR671ML, λ = 671nm, Lasever, Ningbo, China) illuminated the sample and the speckle images were recorded by a 12bit camera (MER-125-30UM, Daheng Imaging, China, 1292×964 pixels, 3.75μm×3.75μm) with an exposure time set between 20 and 40 msec. The DMD (DLC9500P24 0.95VIS) acted as a reflection mirror here. In system calibration, light reflectance from a Lambertian reflection standard was recorded with the exposure time set at 40 msec and a total of 150 images captured at a frame rate of 15 fps. The system characteristics under different levels of light illumination was obtained by varying the intensity attenuator and the reflection ratio of DMD. In flow velocity measurement experiments, Intralipid-2% suspension (scattering coefficient = 1.7 $mm^{-1}$) inside a glass tube (inner diameter 1mm, outer diameter 2mm) is used to simulate blood flow. The flow rate in the glass tube is set by adjusting the driving speed of the fuel injection pump, covering the whole range from 0 to 40 mm/s in this study. A stack of 250 raw speckle images of the dynamic sample was acquired with an exposure time set at 40 msec and a frame rate of 15 fps for each flow speed.

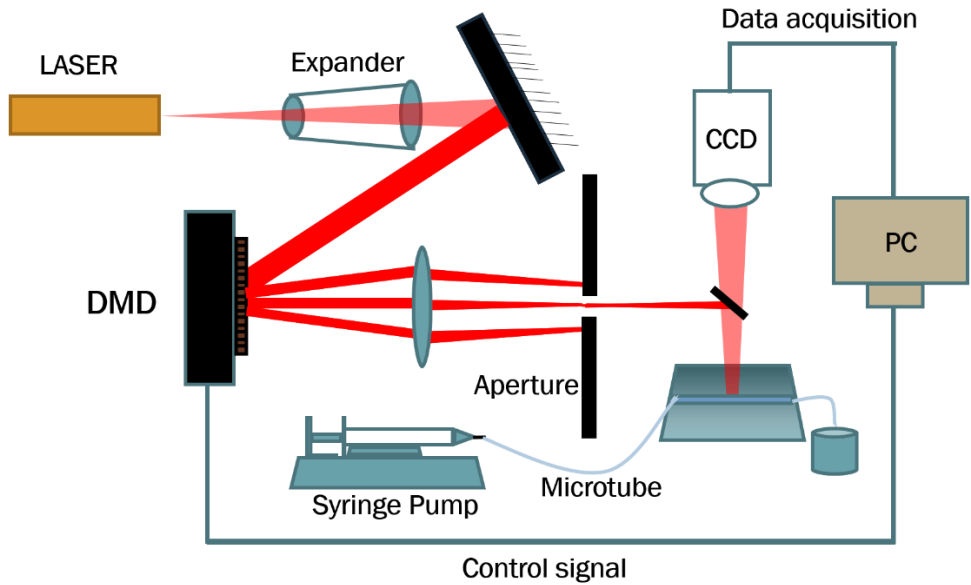

**Fig. 2** Schematic of the experimental setup.

### 3.2 Results of system calibration

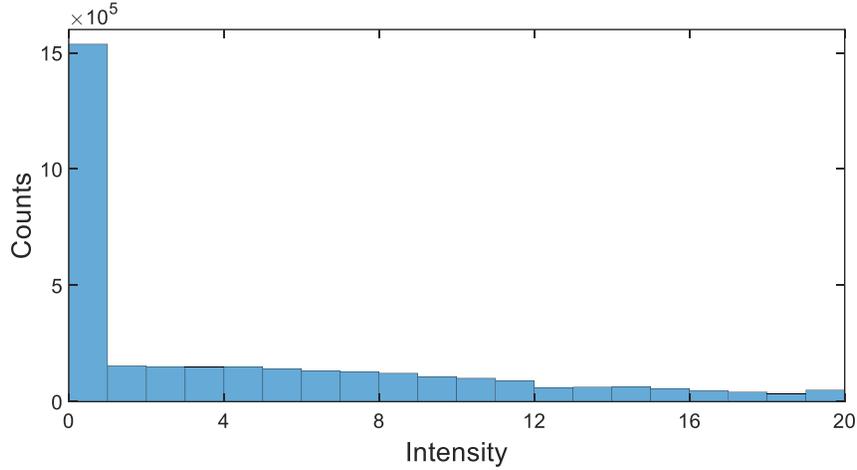

**Fig. 3** Dark current of the camera.

Figure 3 shows the dark current of the camera with a distribution centered at 0. This means that the dark current of the camera has been pre-subtracted and $\langle n_d \rangle$ should be set to 0.

A set of 150 reflectance images from the reflectance standard were then recorded. The coherence factor $\beta$ of the system was then computed with Eq. (19) for different step size $\Delta$ (see Fig. 4). The coherence factor $\beta$ reduces slightly with $\Delta$ owing to the inevitable system instability. The proper system coherence factor is obtained by extrapolating to $\Delta$=0.3 (=0.5×40/67, determined by the exposure time 40 msec and the acquisition time 67 msec), yielding $\beta$ =0.3144.

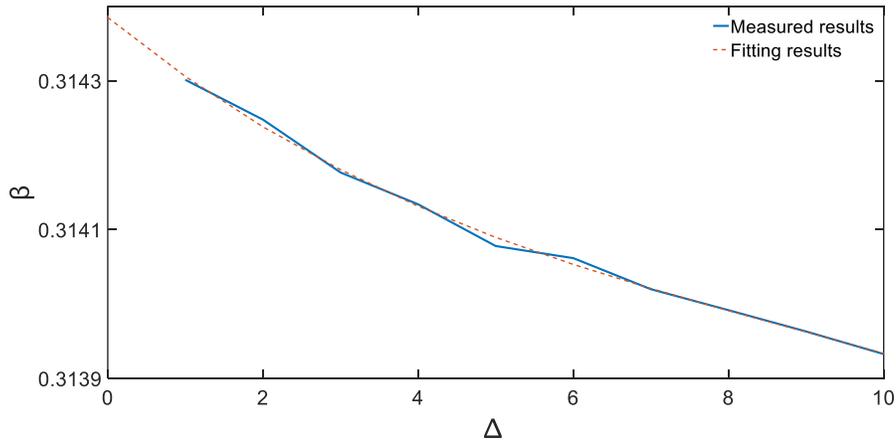

**Fig. 4** The coherence factor $\beta$ reduces with the step size $\Delta$.

By varying the intensity attenuator and the reflection ratio of DMD, multiple sets of 150 reflectance images from the reflectance standard were recorded. The noise variance was computed with Eq. (20) or (21) in the temporal or spatial domain. The noise variance computed with either approach agrees with each other. The noise variance in the spatial domain, however, has lower standard error and is preferred (see Fig. 5). The computed noise variance increases with the step size Δ and the light intensity. The proper noise variance is obtained by extrapolating to Δ=0.3 as well as the determination of β above.

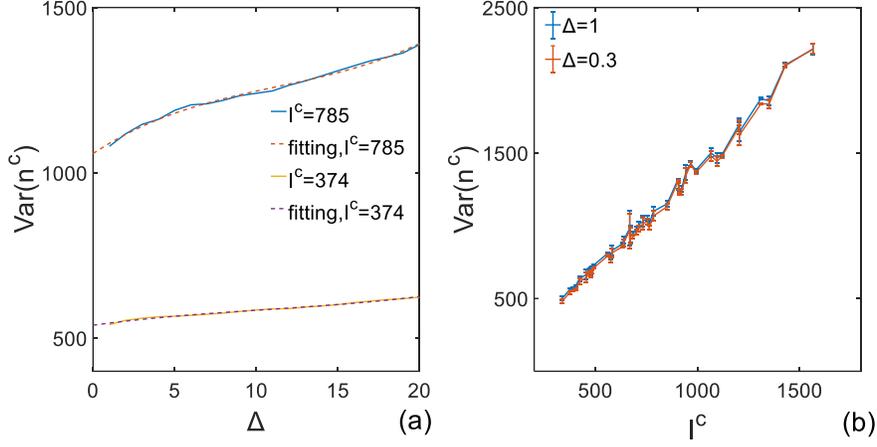

**Fig. 5** Noise variance increases with the step size Δ and the light intensity for (a) two particular intensities, and (b) Δ=0.3 and 1.

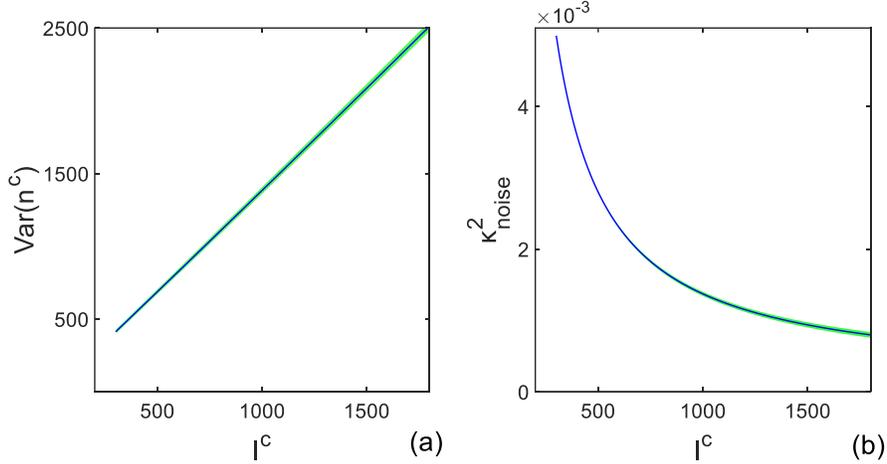

**Fig. 6** Noise characteristics of the imaging system: (a) Noise variance and (b) noise contrast $\kappa_{noise}^2$ versus light intensity extrapolated at Δ=0.3. The shadow represents the error range.

Figure 6 shows the noise characteristics of the imaging system by extrapolating to Δ=0.3. The shadow represents the error range given by the standard deviation computed from five separate sets of measurements. Table 1 displays the noise parameters of $Var(n_i^c)$ by fitting with Eq. (10).

**Table 1** Fitted noise parameters of $Var(n_i^c)$

| $\alpha$ | $\gamma$ | $Var(n_d)$ |
|---|---|---|
| $(1.47\pm0.05)\times10^{-4}$ | $0.89\pm0.03$ | $99\pm2$ |

In previous LSCI experiments, an analog-to-digital conversion factor [9]

$$\gamma' = \frac{\langle I_i^c(x,y) \rangle}{Var(n_i^c) - Var(n_d)} \tag{28}$$

was often used. The value of this factor calculated from the measurement is observed to decrease with the light intensity (see Fig. 7). The assumption of a constant $\gamma'$ is thus not correct, attributed to the nonzero α mainly caused by the light source fluctuations.

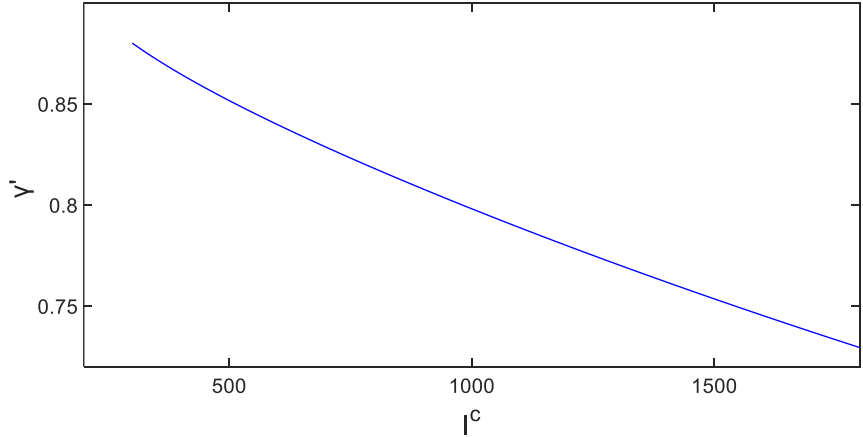

**Fig. 7** The analog-to-digital factor $\gamma'$ decreases with the light intensity.

### *3.3 Results of dynamic sample measurements*

#### 3.3.1 Importance of static scattering removal and noise correction

A stack of 250 images were taken for the flow phantom at each flow speed ranging between 0 and 40 mm/s. The dynamic fraction $\rho$ was computed with Eq. (22). Fig. 8(a) shows the 2D distribution of $\rho$ with an average value of 0.871 over a region of interest (ROI) when flow speed is 0 (Brownian motion alone). The extracted value of $\rho$ stays unchanged when the flow speed increases (see Fig. 8(b)). The non-uniformity of the dynamic fraction is caused by the imperfect glass tube.

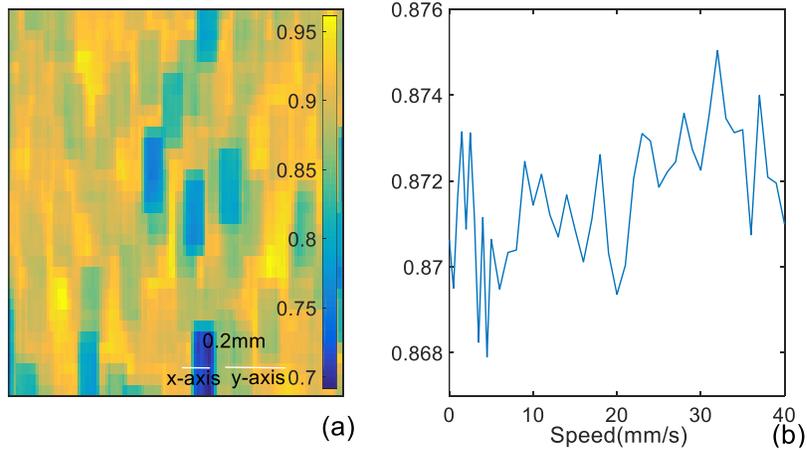

(a)     (b)

**Fig. 8** The dynamic fraction $\rho$ over a ROI. (a) 2D distribution when the flow speed is 0. (b) The average dynamic fraction versus the flow speed.

The noise contrast $\kappa^2_{noise}$ was computed using Eq. (23) or Eq. (24) in the temporal or spatial domain, respectively. Fig. 9 shows the temporal and spatial $\kappa^2_{noise}$ when the flow speed is 0. The temporal $\kappa^2_{noise}$ has much higher spatial resolution than the spatial one. The average temporal $\kappa^2_{noise}$ is $0.00169 \pm 0.00015$ and the average spatial $\kappa^2_{noise}$ is $0.00165 \pm 0.00001$, agreeing with each other.

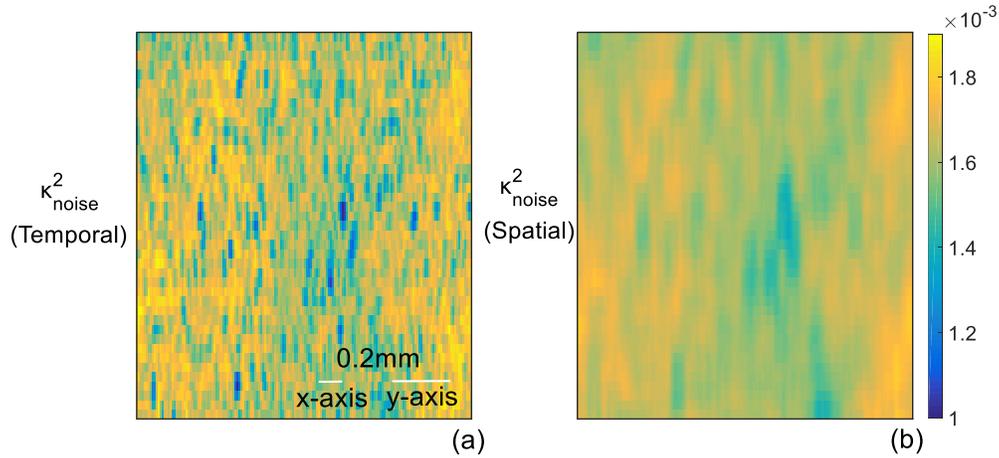

**Fig. 9** $\kappa^2_{noise}$ computed in the (a) temporal and (b) spatial domain.

Figure 10 shows the temporal speckle contrast $K^2$ computed from the data set (original, after noise correction, after both noise correction and static scattering removal yielding $K^2_f$).

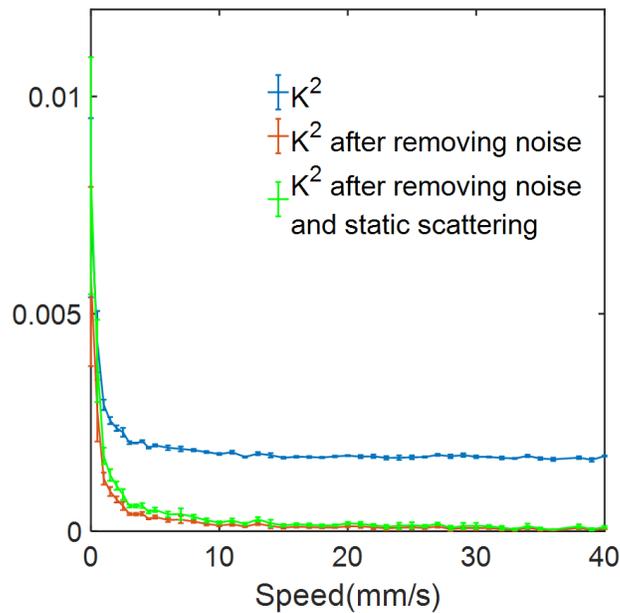

**Fig. 10** Correction of the temporal speckle contrast $K^2$. $K^2$ after both noise correction and static scattering removal yields $K_f^2$.

The flow speed is directly related to the decorrelation time. The inverse decorrelation time $1/\tau_c$ increases with the flow speed and may serve as its proxy. The inverse decorrelation time extracted from fitting Eq. (8) is shown in Fig. 11. The sensitivity of uncorrected $1/\tau_c$ to the flow speed is very poor and loses linearity around 5 mm/s whereas the corrected $1/\tau_c$ shows excellent linearity over the whole range up to 40 mm/s. The corrected one with both noise and static scattering removal also exhibits the least standard deviation.

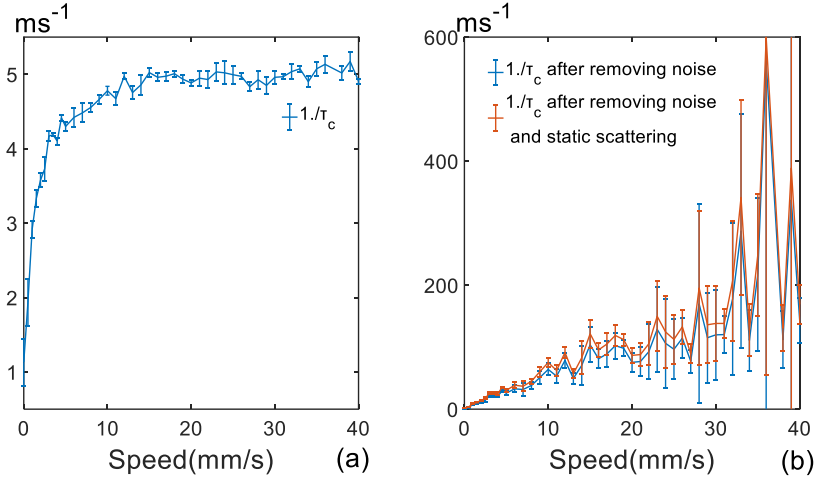

**Fig. 11** The inverse decorrelation time from (a) uncorrected and (b) corrected temporal speckle contrast.

### 3.3.2 Effects of different static scattering

The efficacy of the static scattering removal is then investigated. One part of the glass tube was coated with a scattering layer (dried colloidal suspension) and the same set of the measurements were performed. The region A (ρ=0.83, average light intensity = 670) in the green rectangle is covered by the static scattering layer whereas the region B (ρ=0.88, average light intensity = 810) in the red rectangle is directly exposed (see Fig. 12). Both regions should have identical flow speed.

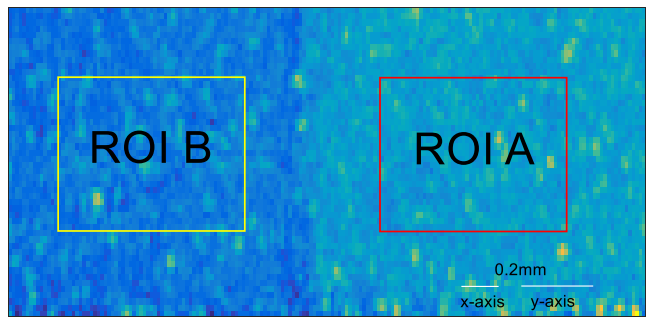

**Fig. 12** ROI A and B (covered with an extra static scattering layer) are imaged.

Figure 13 compares the temporal speckle contrast and the inverse decorrelation time for ROI A and B. The inverse decorrelation time from the uncorrected speckle contrast differs significantly between A and B (see Fig. 13 (a,d)). After noise correction, the agreement between A and B significantly improves although the discrepancy between their recovered $1/\tau_c$ is appreciable (see Fig. 13 (b,e)). With a further static scattering removal, the gap between $1/\tau_c$ for the two regions in (e) almost disappeared (see Fig. 13 (c,f)). The degrade in the performance for faster flow speeds is caused by the loss of SNR at higher speeds. The above results show that different static scattering can be successfully removed to obtain the true flow velocities.

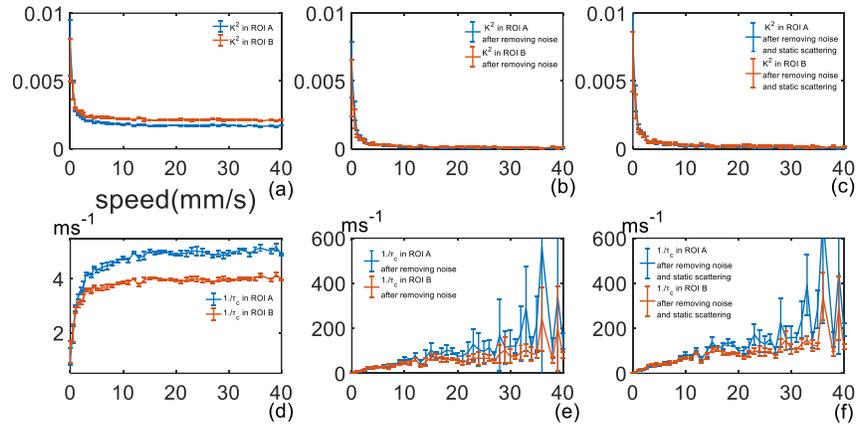

**Fig. 13** (a-c) The temporal speckle contrast (uncorrected, after noise correction, after both noise correction and static speckle removal) for ROI A and ROI B; (d-f) the recovered corresponding inverse decorrelation time.

To further show the agreement of the flow speed in ROI A and B, the error in the recovered $1/\tau_c$ can be directly estimated using the uncertainty in the noise variance. The noise contrast depends on light intensity alone when the imaging system has been specified. At higher speeds, the uncertainty in the noise variance starts to dominate as the flow contrast steadily reduces. Fig. 14 shows the flow speeds in regions A and B indeed agree with each other within the system uncertainty given in Fig. 6 and Table 1.

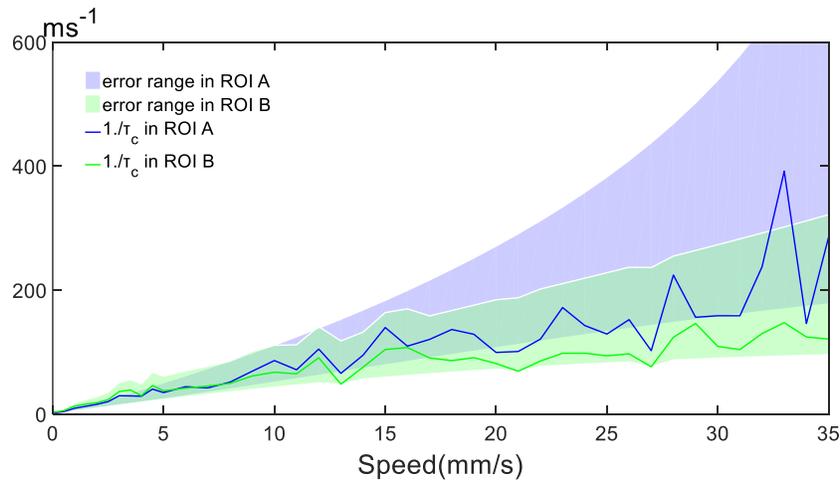

**Fig. 14** The flow speed at region A and B agrees with each other within system uncerntainty.

3.3.3 Temporal speckles vs spatial speckles

The speckle contrast analysis can not only be performed within the temporal domain presented in Sec. 3.3.1 and 3.3.2 but also in the spatial domain. The two different approaches have their own merits.

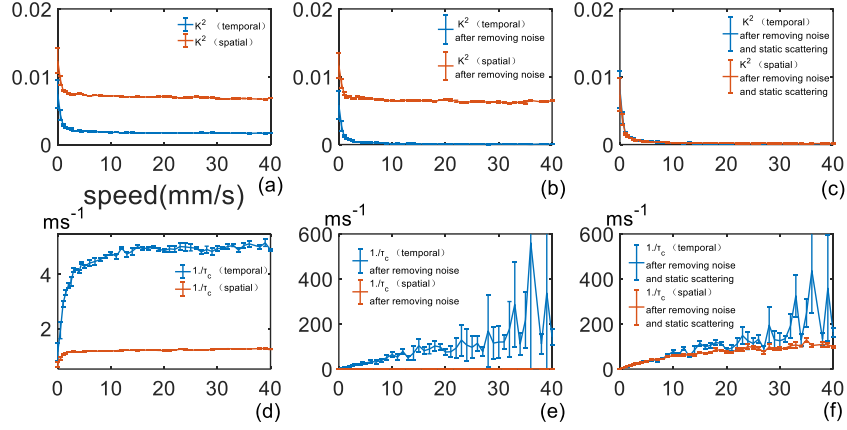

**Fig. 15** (a-c) The temporal and spatial speckle contrast (uncorrected, after noise correction, after both noise correction and static speckle removal) for ROI A; (d-f) the recovered corresponding inverse decorrelation time.

Figure 15 compares the temporal and spatial speckle contrast and the inverse decorrelation time for ROI A. The uncorrected temporal and spatial speckle contrasts and inverse decorrelation times differ significantly caused by the non-ergodic static scattering $\kappa_{ne}^2$ (see Fig. 14 (a,d)). After noise correction, the temporal speckle contrast and inverse decorrelation time performs much better than the spatial counterparts which still retain $\kappa_{ne}^2$ (see Fig. 14 (b,e)). With a further static scattering removal, $K_f^2$ and $1/\tau_c$ in the temporal and spatial domains tend to agree with each other (see Fig. 13 (c,f)). However, a careful examination of the recovered $1/\tau_c$ in (f) reveals their difference. The inverse decorrelation time recovered in the spatial domain shows much less variation yet the linearity range of the temporally recovered $1/\tau_c$ expands to much higher speeds. The latter behavior can be attributed to the difficulty of static scattering removal inside the spatial domain where a subtraction between the measured contrast $K^2$ and $\kappa_{ne}^2 = \beta(N_p)(1-\rho)^2$ is required. At higher speeds, the error in $\kappa_{ne}^2$ dominates and the flow speckle contrast $K_f^2$ computed in the spatial domain fails to obtain the true flow speed.

### 3.3.4 2D flow profile

In addition to the above LSCI analysis of the overall behavior of the flow contrast and inverse decorrelation time versus flow speed, the result of LSCI imaging of a specific 2D region is shown in Fig. 16 (v=3 mm/s). The flow speed is observed to increase closer to the center of the tube. The flow speed cross sectional profile $1/\tau_c$ marked in Fig. 16 (a) fits well by a Newtonian flow profile [22].

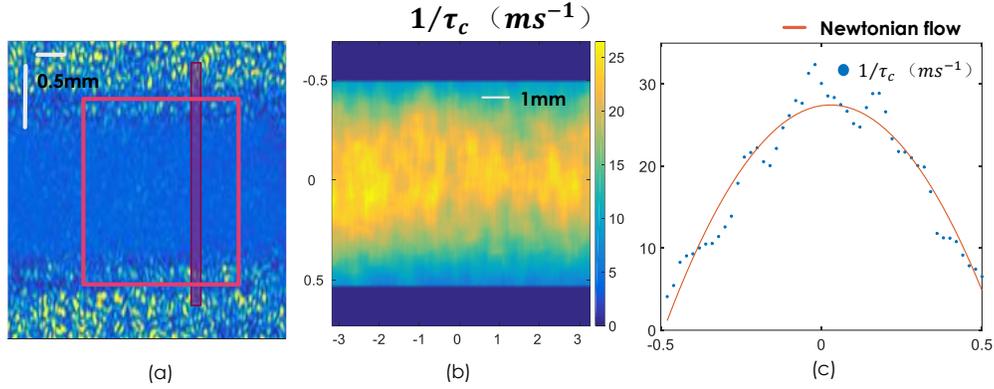

**Fig. 16** (a) ROI selected for analysis. (b) Flow index $1/\tau_c$ for the ROI. (c) The flow speed cross sectional profile $1/\tau_c$ marked in (a) fitted to a Newtonian flow profile ($\propto 0.5^2 - r^2$).

## 4. Discussions

The measured temporal and spatial speckle contrasts for flow imaging are affected by both the presence of static scattering and measurement noise. Their values always differ from each other except for a pure dynamic medium without static scattering. The spatial speckle contrast contains one extra term $\kappa_{ne}^2$ due to the non-ergodic static scattering than the temporal counterpart. Nevertheless, a common true flow speckle contrast $K_f^2$ can be defined in both the temporal and the spatial domains. A complete procedure covering the LSCI system calibration, static scattering removal, and measurement noise estimation and correction to obtain the true flow speckle contrast $K_f^2$ and the flow speed from single-exposure LSCI measurements has been detailed here. The recovered inverse decorrelation time $1/\tau_c$ from $K_f^2$ exhibits excellent linearity against the flow speed over the full span from 0 to 40 mm/s. The true $1/\tau_c$ is obtained regardless of the overlying static scattering layers and the type of measured contrasts (temporal or spatial speckle contrasts).

Comparing speckle contrasts in the temporal and the spatial domain, the latter contains one additional term of $\kappa_{ne}^2$. This fact explains the apparent increase of the spatial speckle contrast with the static scattering [23]. The inverse decorrelation time recovered in the spatial domain shows much less variation yet with a much shorter linearity range than that obtained in the temporal domain. The rapid deterioration of the performance of $K_f^2$ in the spatial domain is caused by the difficulty of static scattering removal which requires a subtraction between the measured spatial speckle contrast $K^2$ and $\kappa_{ne}^2$. At higher speeds, the error in $\kappa_{ne}^2$ dominates and the flow speckle contrast $K_f^2$ can no longer be accurately estimated. This observation is fundamental in selecting the appropriate statistics in analysing the LSCI measurements. A general guideline is that the spatial speckle contrast should be avoided when $K_f^2$ is not larger than the error in $\kappa_{ne}^2$.

A Lorentz velocity distribution model for the moving particles is assumed to relate $K_f^2$ to the flow speed in our study. Different velocity distribution models may be assumed [24]. However, in the typical situation of much longer exposure time compared to the decorrelation time (as in our study), the relation Eq. (8) stills holds other than a trivial pre-factor. The underlying flow speed estimated by $\lambda/2\pi n\tau_c$ compares favorably with the input value where $\lambda$ is the vacuum wavelength of the incident light and *n* is the refractive index of the medium. For example, it yields $2.0\pm0.1$ mm/s when the input flow speed is 5 mm/s and $4.2\pm0.2$ mm/s when the input flow speed is 10 mm/s (see Fig. 15).

Finally, although our study is on the single-exposure laser speckle imaging, the same analysis methodology can be carried over to the multiple-exposure LSCI. The latter gains the advantage over the former when probing the flow velocity distribution. However, the much simpler and faster single-exposure LSCI performs as well as multiple-exposure LSCI when the detailed velocity distribution is not of interest as long as the exposure time is much larger than the decorrelation time and the outlined analysis procedure is followed.

The code for the proposed analysis procedure has been provided in GitHub [25].

## 5. Conclusion

In summary, a systematic and robust laser speckle flow imaging method and procedure has been presented, covering the LSCI system calibration, static scattering removal, and measurement noise estimation and correction to obtain a true flow speckle contrast and the flow speed from single-exposure LSCI measurements. The power of our LSCI analysis recipe has been demonstrated by imaging phantom flow at varying speeds, showing that (1) our recipe greatly enhances the linear sensitivity of the flow index and the linearity covers the full span of flow speeds from 0 to 40 mm/s; and (2) the true flow speed is recovered regardless of the overlying static scattering layers and the type of speckle statistics (temporal or spatial). The difference and merits of the temporal and spatial speckle contrasts have been compared and a guideline for selecting the appropriate statistics for LSCI has been provided. The proposed LSCI analysis framework paves the way to estimate the true flow speed in the wide array of laser speckle contrast imaging applications.

## Funding

Natural Science Foundation of Zhejiang Province (LZ16H180002); National Natural Science Foundation of China (81470081); Wenzhou Municipal Science and Technology; Bureau (ZS2017022); National Science Foundation, USA (1607664).

## Disclosures

The authors declare that there are no conflicts of interest related to this article.